\documentclass[apj]{emulateapj}
\slugcomment{}

\shorttitle{}
\shortauthors{}

\usepackage{color}
\usepackage{amsmath}
\usepackage{booktabs}
\usepackage{enumerate}
\bibpunct{(}{)}{;}{a}{}{,}

\begin{document}

\title{ALMA observations of the protostellar disk around the VeLLO IRAS\,16253--2429}

\author{Tien-Hao Hsieh$^{1}$, Naomi Hirano$^{1}$, Arnaud Belloche$^{2}$, Chin-Fei Lee$^{1}$, Yusuke Aso$^{1}$, and Shih-Ping Lai$^{1,3}$}
\affil{$^{1}$Institute of Astronomy and Astrophysics, Academia Sinica, P.O. Box 23-141, Taipei 106, Taiwan}
\affil{$^{2}$Max-Planck-Institut f\"{u}r Radioastronomie, Auf dem H\"{u}gel 69, 53121 Bonn, Germany}
\affil{$^{3}$Institute of Astronomy, National Tsing Hua University (NTHU), Hsinchu 30013, Taiwan}
\email{thhsieh@asiaa.sinica.edu.tw}

\begin{abstract}
We present ALMA long-baseline observations toward the Class 0 protostar IRAS\,16253-2429 (hereafter IRAS\,16253) with a resolution down to 0\farcs12 ($\sim$15 au).
The 1.3 mm dust continuum emission has a deconvolved Gaussian size of $0\farcs16\times0\farcs07$ (20 au $\times$ 8.8 au), likely tracing an inclined dusty disk.
Interestingly, the position of the 1.38 mm emission is offset from that of the 0.87 mm emission along the disk minor axis. 
Such an offset may come from a torus-like disk with very different optical depths between these two wavelengths.
Furthermore, through CO ($2-1$) and C$^{18}$O ($2-1$) observations, we study rotation and infall motions in this disk-envelope system and infer the presence of a Keplerian disk with a radius of $8-32$ au.
This result suggests that the disk could have formed by directly evolving from a first core, 
because IRAS\,16253 is too young to gradually grow a disk to such a size considering the low rotation rate of its envelope.
In addition, we find a quadruple pattern in the CO emission at low velocity, which may originate from CO freeze out at the disk/envelope midplane.
This suggests that the ``cold disk'' may appear in the early stage, implying a chemical evolution for the disk around this proto-brown dwarf (or very low-mass protostar) different from that of low-mass stars.
\end{abstract}


\section{INTRODUCTION}
Rotationally supported disks or Keplerian disks are commonly seen in young stellar objects (YSO) at the Class II stage \citep{wi11,be13}.
Only a handful of disks have been kinematically identified at earlier evolutionary stages, at the Class I stage \citep{lo08,ta12,br13,ch14,ha14,li14,ye14,as15,le16}, and even rarer at the Class 0 stage \citep{to12a,mu13,oh14,le17b,le18,as17}.
Because disks are believed to grow rapidly after the start of the core collapse \citep{te84,wi11}, star-disk systems in an embedded phase are valuable for understanding disk formation especially for that at the Class 0 stage. 
In addition, it is also unclear whether the disk formation channel and evolution depend on the stellar mass, for instance whether disks form and evolve differently in brown dwarfs compared to higher-mass objects \citep{ri14,te16}.

IRAS\,16253--2429 (hereafter IRAS\,16253) was first discovered as a Class 0 source by \citet{kh04} in the $\rho$ Oph star forming region (\citealp[$d=125~{\rm pc}$,][]{ev09}). Later, it was classified as a Very Low Luminosity Object (VeLLO) owing to its internal luminosity of $\approx$0.09 $L_{\odot}$ \citep{du08}.
Such a low luminosity implies that IRAS\,16253 is an extremely young Class 0 protostar, a very-low mass protostar, or a combination of both \citep{du14}. Using the deuterium fraction and outflow opening angle as evolutionary indicators, \citet{hs15,hs17} suggest that IRAS\,16253 is a young Class 0 object.
Assuming that the infalling motions derived from C$^{18}$O observations to be free-fall, \citet{ye17} estimate its central mass to be $0.03~M_\odot$. In addition, the envelope mass has been estimated to be $0.2-0.8~M_\odot$ \citep{ba10,st06,en08,to12b}.
These results, together with the low outflow force \citep{hs16}, suggest that IRAS\,16253 may form a brown dwarf or very low-mass star depending on its future accretion.
Although this substellar object unlikely hosts a sizeable protostellar disk, the integrity of its bipolar outflow implies the existence of a disk.
In addition, \citet{hs18} found that IRAS\,16253 has experienced a past accretion burst based on the outward shift of the CO snow line. This is believed to originate from a gravitationally instability of the disk \citep{vo13,vo15}.

CO and its isotopologues have been used to probe the kinematics of disks due to the brightness of their rotational transitions at submillimeter/millimeter wavelengths.
With high-resolution ALMA observations, multiple CO isotopologues further provide a powerful diagnostic of their density and temperature structures \citep{no16,sc16,wa17}.
However, recent works find that CO could be depleted near the midplanes of the Class II disks within a few hundred au \citep{qi11,zh17,hu17,pi18}.
This could be explained by freeze out of CO onto the dust grains or conversion of CO into less volatile molecules \citep{ai15,va17}

In this paper, we present new ALMA observations of CO, C$^{18}$O, and 217 GHz continuum in the proto-brown dwarf candidate IRAS\,16253.
We aim to search for a disk around this unique source and study its physical and chemical properties.
The observations and results are described in Section 2 and Section 3, respectively. In Section 4, we detail our analyses and models of the continuum and line emission. Finally, the discussion and summary are given in Sections 5 and 6, respectively.

\begin{figure}
\includegraphics[width=0.5\textwidth]{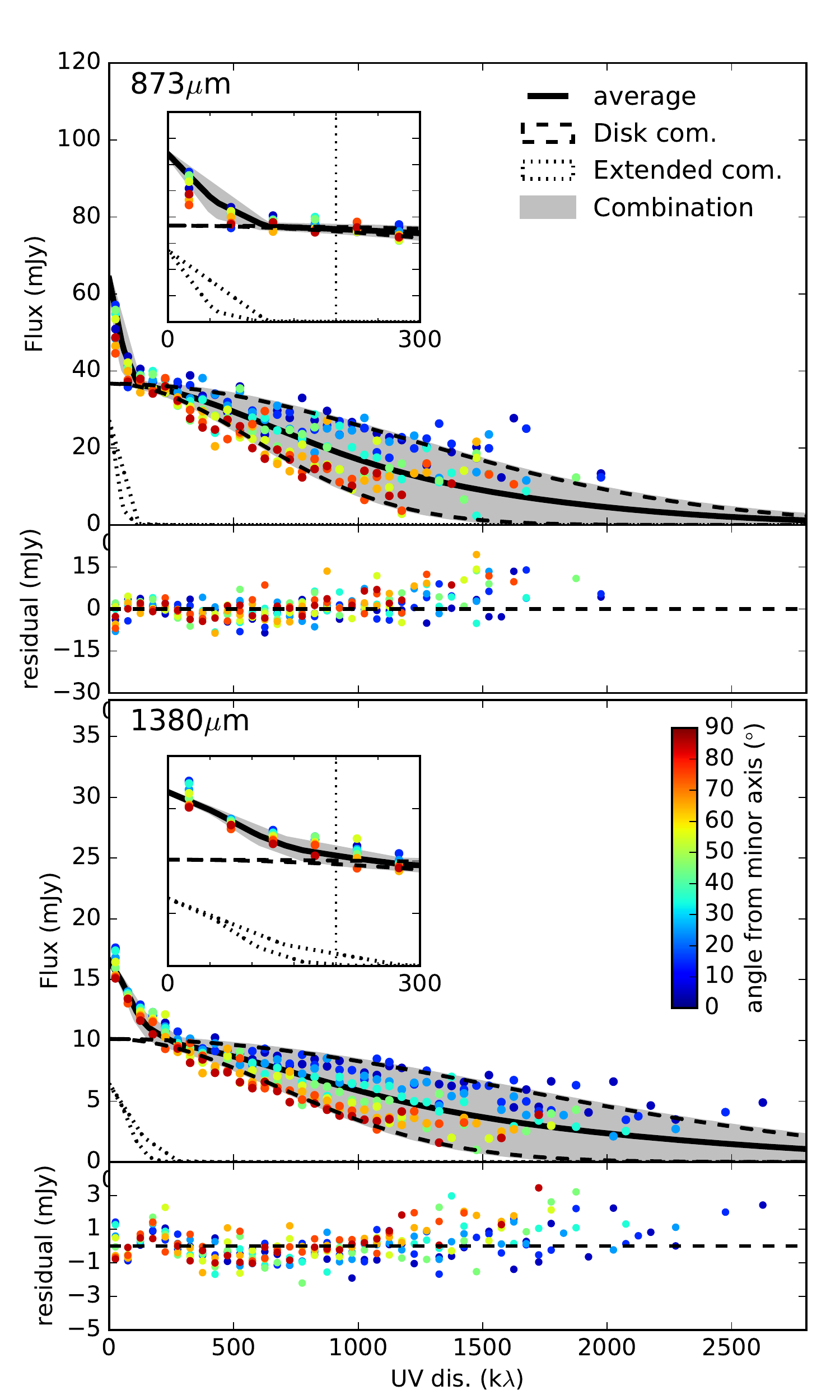}
\caption{$uv$-distance versus amplitude for 0.87 mm (top) and 1.38 mm (bottom) data with the residual from the best-fit. The color indicates the angles relative to the minor axis in $uv$-space, i.e., blue points close to the minor axis and red points close to the major axis. 
The grey area shows the best-fit two-component model, i.e. amplitudes along the minor and major axes as the upper and lower boundaries, and the solid line represents their average (Table \ref{tab:gau}).
The dashed and dotted curves represent the disk and extended components, respectively. The subplot in each panel shows a zoom-in of $0-300~{\rm k}\lambda$ with the vertical dashed line indicating $200~{\rm k}\lambda$.}
\label{fig:gau}
\end{figure}

\section{Observations}
We used ALMA to simultaneously observe CO ($2-1$), C$^{18}$O ($2-1$), and dust continuum emission at 217 GHz toward IRAS\,16253 ($\alpha$=16$^{\rm h}$28$^{\rm m}$21\fs6, $\delta$=$-$24\arcdeg36\arcmin23\farcs4) on 2017 Aug 18th (Cycle 4 project, 2016.1.00598.S) with 42 available antennas. The total observing time was 48 min and the on-source time was $\approx$20 min.
The data were obtained in configuration of C40-7 with a projected baseline range of 11 to 2629 k$\lambda$, resulting in a spatial resolution of $0\farcs11\times0\farcs08$ with uniform weighting for the continuum and of $0\farcs14\times0\farcs11$ with natural weighting for CO ($2-1$).
The C$^{18}$O ($2-1$) data were combined with the ALMA cycle 2 data from \citet{ye17} to increase the sensitivity. 
However, although the $uv$-coverages overlap in these two data sets, the non-uniform sampling produces a clean beam featuring a summation of distinct large and small beam.
After iterations, we selected data with a $uv$-distance shorter than 600 k$\lambda$, resulting in a relatively Gaussian-like beam with a size of $0\farcs40$ by $0\farcs37$.
The continuum bandwidth was 1840 MHz, centered at 217 GHz (1.38 mm).
The channel width was 122 kHz (0.16 km s$^{-1}$) for CO ($2-1$) and 61 kHz (0.08 km s$^{-1}$) for C$^{18}$O ($2-1$).
The rms noise level is 0.066 mJy beam$^{-1}$ for the continuum map, and 5.6 and 4.9 mJy beam$^{-1}$ for the CO and C$^{18}$O maps with a channel width of 0.16 km s$^{-1}$, respectively.
The bandpass, flux, and phase calibrators were J1517-2422, J1733-1304, and J1625-2527, respectively. 
A check source, J1626-2951, was observed for 5 scans spread about uniformly between the 26 scans on IRAS\,16253. Self-calibration is not applied in order to maintain astrometric information of the source.

In order to compare with the 1.38 mm continuum emission, we obtained 0.87 mm data from the ALMA archive (2015.1.00741.S, PI: L. Looney). 
These data are obtained with a compact configuration and an extended configuration. The $uv$ range from 15 to 2031 k$\lambda$ is comparable to our 1.38 mm data. The on-source time is 30 and 60 sec for the compact and extended configurations, respectively. The bandwidth is $4\times1840~{\rm MHz}$.
The bandpass, flux, phase calibrators, and check source 
are J1517-2422, J1733-1304, J1625-2527, and J1627-2426, respectively for the compact configuration and 
J1517-2422, J1517-2422, J1625-2527, and J1633-2557, respectively, for the extended configuration.

\begin{figure*}
\includegraphics[width=1.02\textwidth]{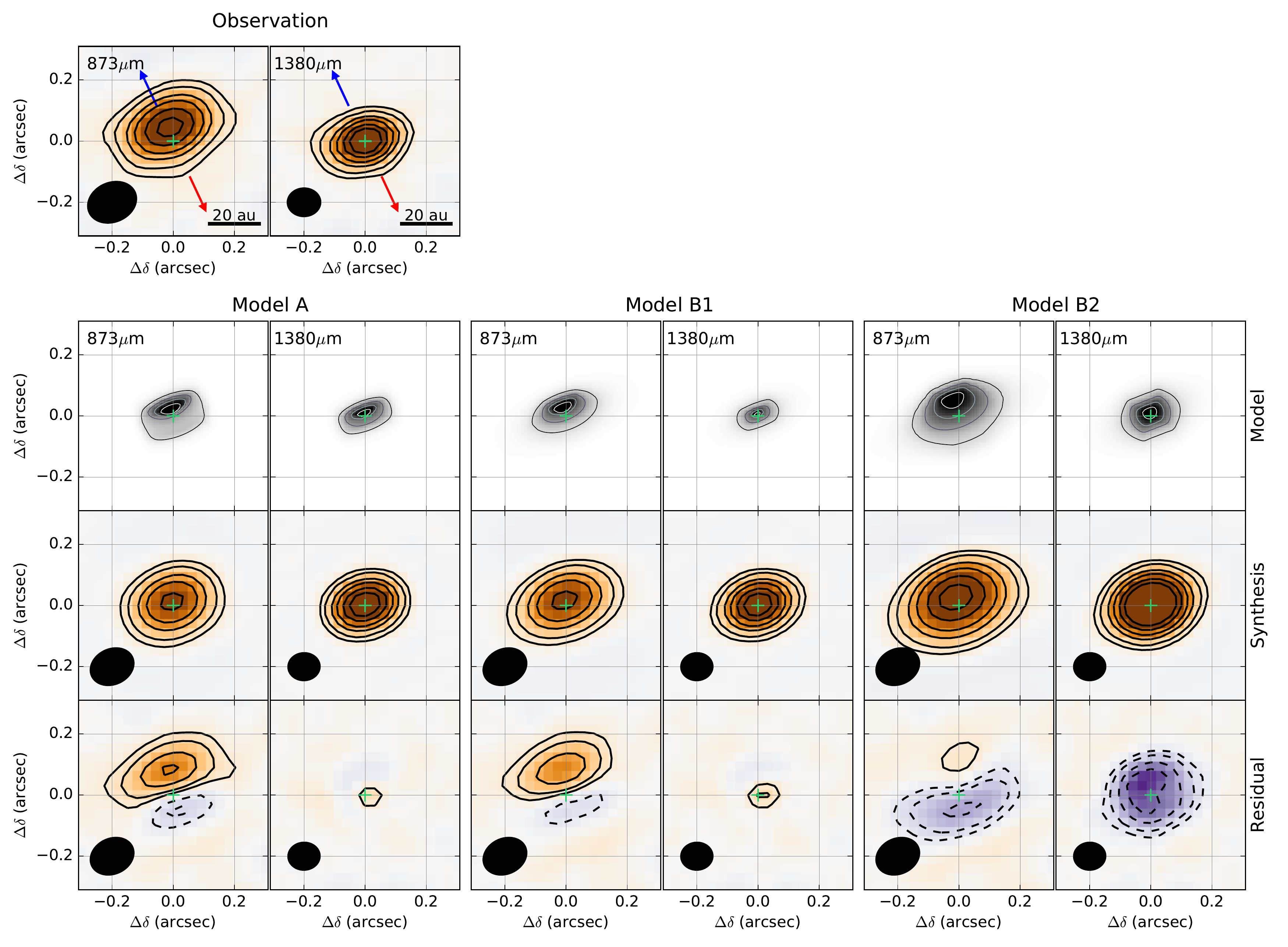}
\caption{Observed and modeled images at 0.87 mm and 1.38 mm (left and right columns of all panels) using only visibilities at uv-distances beyond 200 k$\lambda$.
The top panel shows the observed images with the contour levels at 5, 10, 20, 30, 45, 60$\sigma$. The rms noise level $\sigma$ is 0.34 mJy beam$^{-1}$ (165 mK) at 0.87 mm and 0.066 mJy beam$^{-1}$ (168 mK) at 1.38 mm. The bottom panels are the three models (see the text). The top row shows the model with contours at levels of 20\%, 40\%, 60\%, 80\% of the peak flux, the second row shows the images from synthetic observations with the same contour levels, and the bottom row shows the residuals between observations and models. The green plus sign indicates the source center from the Gaussian fitting at 1.38 mm.}
\label{fig:con}
\end{figure*}

\section{Results}
\subsection{continuum emission at 0.87 mm and 1.38 mm}
The continuum emission at 1.38 mm and 0.87 mm is found to contain two components (Figure \ref{fig:gau}).
We employ a double Gaussian fit to the observed amplitude but not phase, and we find a break point at a $uv$-distance of $\approx$200 k$\lambda$ separating the extended and compact components.
The fit of amplitude could introduce a bias toward positive values at high uv distances when the S/N is low, but it does not significantly affect the location of the break point at the low uv distance.
Although the extended component shows ambiguous fitting results, the fitted compact component is consistent in position angle and aspect ratio for both wavelengths (Table \ref{tab:gau}).
The minor axis aligns well with the outflow orientation (\citealp[$\sim$20\arcdeg,][]{hs17}) and the aspect ratio is consistent with the inclination angle derived from the outflow (\citealp[$60\arcdeg-65\arcdeg$, with $0^\circ$ for pole-on]{ye17}).
Therefore, the compact component likely traces a disk.

Figure \ref{fig:con} shows the images made with visibilities at uv-distances beyond 200 k$\lambda$ at 0.87 mm and 1.38 mm (Figure \ref{fig:con}).
These images indicate that IRAS\,16253 remains a single source at $\approx0\farcs1$ resolution. This result is contrary to our previous prediction that the directional variability of the protostellar jet comes from a rotating binary system with a separation of $0\farcs55$ \citep{hs16}.
Another mechanism \citep{ra09,of17,le17} is needed to explain the directional change of the outflow axis.

Although the fitting results of the compact component are similar in position angle and aspect ratio at the two wavelengths, the central positions are offset by 46 milliarcsecs (mas) (Figure \ref{fig:con}).
To assess the accuracy of the calibration, we check the check sources, J1626-2951 for 1.3 mm and J1633-2557 for 0.87 mm, and find that they are consistent with the referenced positions within 10 mas and 5 mas, respectively.
However, IRAS\,16253 was observed by only one scan with the extended configuration for 0.87 mm, and the check source was not taken in this interval between the scans of the phase calibrator.
Thus, the phase calibration is generally good, but we are not able to completely exclude the possibility that the offset results from an inaccurate phase calibration of the 0.87 mm data.

\tabletypesize{\scriptsize}
\begin{deluxetable*}{ccccccccc}
\tabletypesize{\tiny}
\tablecaption{\bf Results of UV Gaussian fitting}
\tablehead{ 
\colhead{}
& \multicolumn{3}{c}{Extended component}
& \multicolumn{3}{c}{Disk component}
\\ 
\cmidrule(lr){2-4} \cmidrule(lr){5-7}
\colhead{wavelength}		
& \colhead{Flux}		
& \colhead{Size}
& \colhead{P.A.}	
& \colhead{Flux}	
& \colhead{Size}
& \colhead{P.A.}	
\\
\colhead{}		
& \colhead{mJy}		
& \colhead{mas}
& \colhead{$^\circ$}	
& \colhead{mJy}		
& \colhead{mas}
& \colhead{$^\circ$}	
}
\startdata 
0.87 mm	& 27.7$\pm$3.6	& 4390$\pm$330$\times$2830$\pm$210	& 22.9$\pm$6.0	& 36.7$\pm$1.9	& 190$\pm$4$\times$91$\pm$3	& 24.9$\pm$1.5\\
1.38 mm	& 6.4	$\pm$0.4		& 1660$\pm$50$\times$1170$\pm$30	& 9.9$\pm$2.7		& 10.1$\pm$0.4	& 161$\pm$2$\times$69$\pm$2	& 21.1$\pm$0.9
\enddata
\label{tab:gau}
\end{deluxetable*}

Here we propose two possibilities to explain this offset: (1) the proper motion of IRAS\,16253 (from 2016 Aug to 2017 Aug) and (2) the different optical depths between both wavelengths.
For the first case, the projected velocity would have to be 49 mas yr$^{-1}$ ($\sim$29 km s$^{-1}$) which is much larger than that of the sources in the $\rho$ Oph region, $\lesssim$10 mas yr$^{-1}$ \citep{du17}. 
However, it is noteworthy that a scenario to form brown dwarfs is the ejection from a relatively massive system that can cause a high source velocity \citep{ba02,ba12}. 
The second possibility is hinted by the offset orientation which is almost along the outflow axis; the 0.87 mm continuum emission may trace the upper layer of the inclined disk due to the high optical depth.

\subsection{$^{12}$CO and C$^{18}$O ($2-1$)}
\begin{figure*}
\includegraphics[width=.91\textwidth,angle=0]{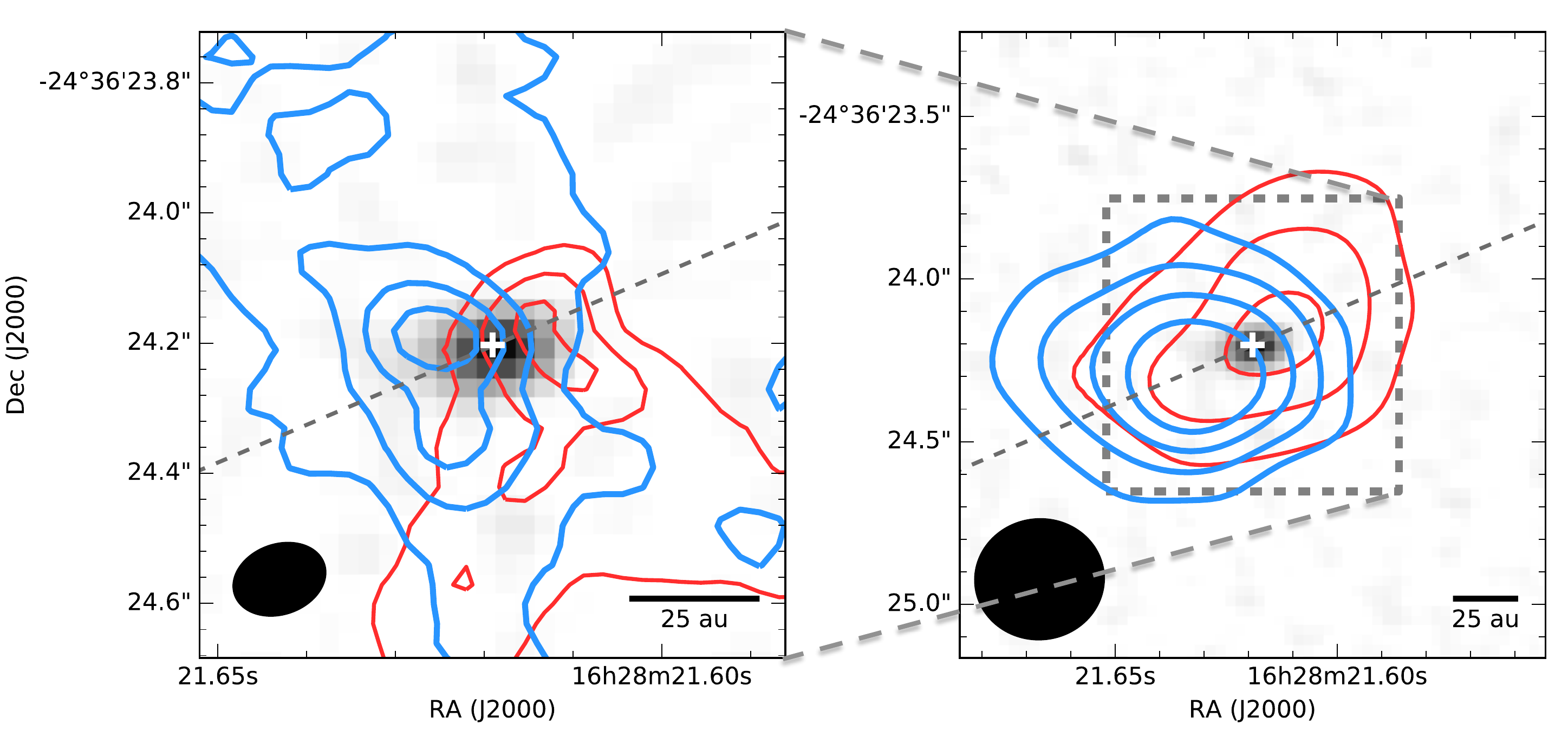}
\caption{({\it left}) CO ($2-1$) integrated intensity maps (contours) overlaid on the 1.38 mm continuum image. The contour levels are 5, 10, 15, and 20$\sigma$ with a rms noise level $\sigma$ of 3.0 mJy beam$^{-1}$ km s$^{-1}$ for both blue- (0.4-2.6 km s$^{-1}$) and red-shifted contours (5.8-8.0 km s$^{-1}$). 
The white plus sign indicates the continuum source position, and the grey dashed line shows its major axis, the PV cut for Figure \ref{fig:pv}.
({\it right}) Same as the left panel but for C$^{18}$O ($2-1$). The rms noise levels are 2.7 and 2.0 mJy beam$^{-1}$ km s$^{-1}$ for the blue- and red-shifted lobes, respectively. The integrated velocity ranges are $1.8-3.3$ km s$^{-1}$ and $4.9-5.7$ km s$^{-1}$ for blue- and red-shifted contours.}
\label{fig:mom}
\end{figure*}

Figure \ref{fig:mom} shows the CO (left) and C$^{18}$O (right) integrated intensity maps, revealing the rotation motion around the outflow axis \citep{hs17}.
The CO map has a much better S/N ratio such that it allows us to trace the high-velocity small-scale structures.
The C$^{18}$O map is combined with the low-resolution data \citep{ye17} and a $uv$ taper of $<$600 k$\lambda$ is applied.
Thus, it has a lower spatial resolution and is dominated by the outer low-velocity region.

\section{Analysis and Discussion}
\subsection{Models of the continuum images}
\label{sec:con}
To explain the offset between the dust continuum peaks at 1.38 mm and 0.87 mm (Figure \ref{fig:con}), we model the emission using the Monte Carlo radiative transfer code RADMC-3D\footnote{http://www.ita.uni-heidelberg.de/~dullemond/software/radmc-3d/} \citep{du12}. 
The dust opacity $\kappa$ as a function of wavelength is constructed using DIANA Opacity Tool\footnote{http://dianaproject.wp.st-andrews.ac.uk/data-results-downloads/fortran-package/} \citep{wa16}. The grain size distribution is assumed to follow $n(a)da\propto a^{-a_{\rm pow}}$ with a maximum size of $a_{\rm max}$.
We construct model grids (Sections \ref{sec:modA} and \ref{sec:modB}) to perform a $\chi^2$ fitting 
to the visibilities at uv-distances beyond 200 k$\lambda$.
The $\chi^2$ is calculated in the complex space (i.e. Eq. 2 in \citealp[][]{as17}) with the modeled values at the uv points covered by the observations, for which the modeled visibilities are computed from the synthetic images using vis\_sample\footnote{The vis\_sample Python package is publicly available at https://github.com/AstroChem/vis\_sample or in the Anaconda Cloud at https://anaconda.org/rloomis/vis\_sample.}

\begin{deluxetable*}{ccccccccc}
\tabletypesize{\tiny}
\tablecaption{Parameters of disk models}
\tablehead{ 
\colhead{Model}	
& \colhead{$L_{\rm star}$}	
& \colhead{$M_{\rm disk}$}		
& \colhead{$a_{\rm pow}$}
& \colhead{$a_{\rm max}$}
& \colhead{disk type}
& \colhead{density pars.}
& \colhead{$\chi^2_{\rm r}$}	
& \colhead{offset}
\\
\colhead{}		
& \colhead{10$^{-2} $L$_\odot$}	
& \colhead{10$^{-3} $M$_\odot$}		
& \colhead{}
& \colhead{$\mu$m}
& \colhead{}
& \colhead{}
& \colhead{}
& \colhead{mas}
}
\startdata 
Model A	& 2.7$\pm$0.3	& 1.6$\pm$0.1	& 3.0\tablenotemark{a}	& 150$\pm$50	& flared	& $\theta_{\rm flared}=18\pm1\arcdeg,~R_{\rm disk}=9\pm1~{\rm au}$	& 1.70	& 7\\
Model B1	& 2.1$\pm$0.2	& 1.5$\pm$0.1 	& 3.0\tablenotemark{a}	& 150$\pm$50	& thick	& $R_{\rm disk}=30\pm2~{\rm au},H_{\rm t}=3.0\pm0.2~{\rm au},~R_{\rm t}=7\pm1~{\rm au}$ 	& 1.59	& 12\\
Model B2\tablenotemark{b}	& 2.1	& 2.5	& 3.0		& 150	& thick	& $R_{\rm disk}=30~{\rm au},H_{\rm t}=5~{\rm au},~R_{\rm t}=7~{\rm au}$	& 6.09	& 20
\enddata
\tablecomments{\bf The error of the fitted parameters are calculated with the $\chi^2$ distribution in a confidence level of 99.9\%.}
\tablenotetext{a}{\bf The opacity spectral index reaches the lower limit we set in the space of free parameters such that no error is provided.}
\tablenotetext{b}{Model B2 uses the same parameters as Model B1 except for $H_{\rm t}=5~{\rm au}$. The conservation of density at disk midplane ($\rho_0$ in Eq. \ref{eq:modB}) results in different total disk masses.}
\label{tab:mod}
\end{deluxetable*}

\subsubsection{Model A - flared-disk model}
\label{sec:modA}
Our model A assumes a flared-disk density structure defined as
\begin{equation}
\rho(r,z)=\frac{\Sigma(r)}{\sqrt{2\pi}H(r)} \exp{[-\frac{1}{2}(\frac{z}{H(r)})^2]}
\label{eq:den}
\end{equation}
with
\begin{equation}
\Sigma(r)=\Sigma_0(\frac{r}{r_0})^{-1}\exp{[-\frac{r}{R_{\rm disk}}]}
\end{equation}
and
\begin{equation}
H(r)=H_0(\frac{r}{r_0})^{1.3} {\rm~and}~H_0 = r_0\tan(\theta_{\rm flared}),
\end{equation}
\citep{ha15}
where $r$ and $z$ are the cylindrical coordinates, $\Sigma_0$ is the disk surface density, $r_0$ is the reference radius 25 au, $R_{\rm disk}$ is the disk radius, and $H_0$ is the scale height at $r_0$ determined by $\theta_{\rm flared}$.

We take six free parameters including the protostellar luminosity ($L_{\rm star}$), $M_{\rm disk}$ (total mass of the disk scaled by $\Sigma_0$), $R_{\rm disk}$, $\theta_{\rm flared}$, $a_{\rm pow}$, and $a_{\rm max}$. The central position, position angle, and inclination angle are fixed based on the 1.38 mm Gaussian fitting; an inclination angle of $65\arcdeg$ is estimated from the Gaussian aspect ratio (Table \ref{tab:gau}).
The best-fit parameters are listed in Table \ref{tab:mod} and the images are shown in Figure \ref{fig:con}.
This fitting converged to an extreme case with $a_{\rm max}=150~\mu$m and $a_{\rm pow}=3.0$ (boundary in the model grid) when it has the largest opacity ratio ($\kappa_{0.87 \rm mm}/\kappa_{1.38 \rm mm}\approx5.5$) between 0.87 mm and 1.38 mm (Appendix A.).
This result is predictable given the large offset between the 0.87 mm and 1.38 mm emission peaks. However, even with this large ratio, the offset (7 mas) in this model is still much smaller than the observed one (46 mas).

\subsubsection{Model B1/B2 - Thick-disk model}
\label{sec:modB}
To reproduce a larger offset, we adopted the torus-like disk model from \citet{le17a,le17b} by adding an exponentially decreasing scale height beyond a radius $R_{\rm t}$, 
\begin{equation}
\label{eq:modB}
\rho(r,z)=\rho_0(\frac{r}{R_{\rm t}})^{-2}\exp{[-\frac{1}{2}(\frac{z}{H(r)})^2]}
\end{equation}
with
\begin{equation}
H(r)=\begin{cases}
    H_0(\frac{r}{R_{\rm t}})^{1.3}, & {\rm if}~r<R_{\rm t}.\\
    H_0\exp{[-(\frac{r-R_{\rm t}}{R_{\rm disk} -R_{\rm t}})^2]}, & {\rm if}~r>R_{\rm t}.
  \end{cases}
\end{equation}
In comparison to the flared disk model, this model includes an additional free parameter, $R_{\rm t}$, to determine the location of the maximum scale height.
The best-fit model (Model B1) is shown in Figure \ref{fig:con}, and the corresponding parameters are listed in Table \ref{tab:mod}.
Compared to Model A, this model has a larger offset 12 mas and better reproduces the elongated shape especially at 0.87 mm because the emission from the lower surface is highly attenuated in the torus-like structure.
However, the offset in the synthetic images is still much smaller than the observed one.

In order to reproduce a larger offset, we defined another model, Model B2, like Model B1 but with $H_{\rm t}=5~{\rm au}$.
As a result, Model B2 has a larger offset of 20 mas than Model B1, but it is still smaller than the observed offset of 46 mas.
Besides, this model has much larger flux densities than the observation especially at 1.38 mm. 

Although Model B2/B1 cannot fit the observations well, they suggest that the offset distance can be affected by the disk density structure.
Future multi-wavelength observations are required to perform a better model which should consider (1) an accurate disk center measured from optically thin emission at long wavelengths, (2) possible external heating to compute accurate flux densities, and (3) different dust components if dust settling has started.

\subsection{PV-diagram and dynamic model}
\label{sec:pv}
\begin{figure*}
\includegraphics[width=0.87\textwidth]{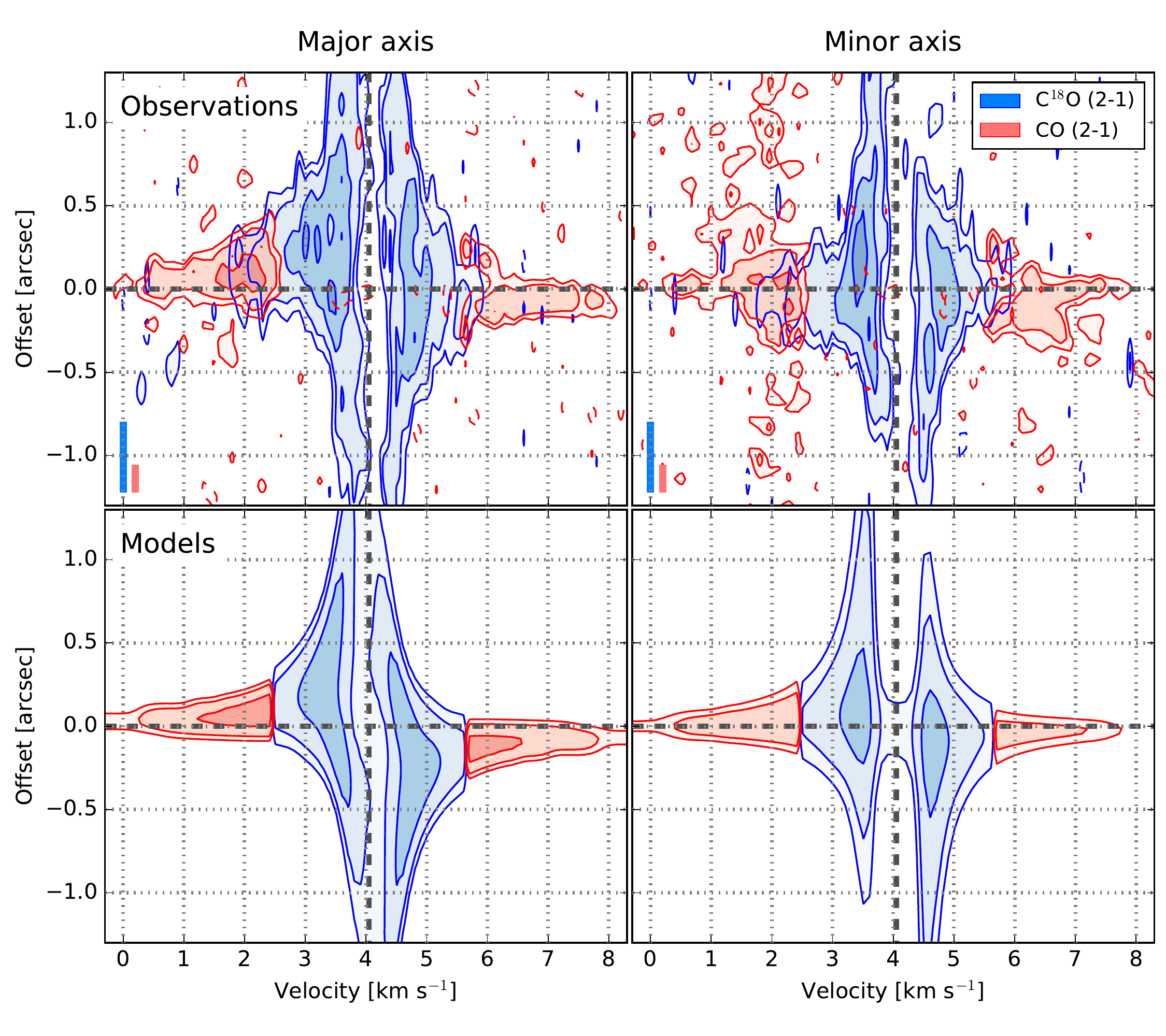}
\caption{PV diagrams of CO emission (red) and C$^{18}$O emission (blue) along the major and minor axes (Figure \ref{fig:mom}). The contours correspond to 3, 5, 10, and 15$\sigma$, where $\sigma$ is 2.9 mJy beam$^{-1}$ for CO and 4.8 mJy beam$^{-1}$ for C$^{18}$O. 
The bottom panel shows the best-fit model with the same contour levels, and it includes only the velocity ranges used in the fitting $\mid V-V_{\rm lsr} \mid<1.6$ km s$^{-1}$ for C$^{18}$O and $\mid V-V_{\rm lsr} \mid>1.6$ km s$^{-1}$ for CO.
The vertical and horizontal dashed lines indicate the systemic velocity and source position.
The bars in the bottom left corner represent the beam size in the same color as the contours.}
\label{fig:pv}
\end{figure*}

Figure \ref{fig:pv} shows the PV diagrams of CO and C$^{18}$O along the major and minor axes centered on the 1.38 mm continuum source.
The emission of both CO and C$^{18}$O emission is likely attenuated by the foreground cloud core near the systemic velocity, especially for CO.
On the other hand, due to the low abundance, C$^{18}$O emission is faint in the high-velocity region where CO is bright.

We model the PV diagrams assuming a rotating infalling envelope with conservation of angular momentum.
We use the radiative transfer code from \citet{le14} to perform the PV model given temperature and density structures under local thermodynamic equilibrium (LTE) conditions.

We use the flared density structure (i.e., Eq. \ref{eq:den}) with an additional free parameter $p$ adjusting its radial distribution,
\begin{equation}
\rho(r,z)=\rho_0 (\frac{r}{r_0})^p\exp{[-\frac{1}{2}(\frac{z}{H(r)})^2]}.
\end{equation}
and the temperature profile is assumed to be
\begin{equation}
T(r,z)=T_0{(r/r_0)^{-0.4}}.
\end{equation}
where $\rho_0$ and $T_0$ scale the gas density and temperature \citep{le14,ye17}.
We take the CO abundance relative to H$_2$ as $X_{\rm CO}=5\times10^{-4}$ and the C$^{18}$O isotopic ratio of $X_{\rm CO}/X_{\rm C^{18}O}=560$ \citep{wi94}.
Because $\rho_0$ and $T_0$ determine the brightness scale and in turn are degenerate, we assume $T_0$ to be 70 K, which should not affect the fitting result of the dynamical structure.
As a result, three free parameters, $\rho_0$, $p$, and $\theta_{\rm flared}$, are used to determine the physical conditions.
The rotation and radial velocities are assumed to follow the conservation of angular momentum and free fall as
\begin{equation}
V_{\rm rot}=V_{\rm rot,0}(\frac{r}{r_0})^{-1.0}
\end{equation}
and
\begin{equation}
V_{\rm infall}=V_{\rm infall,0}(\frac{r}{r_0})^{-0.5}
\label{eq:infall}
\end{equation}
respectively.
This adds two free parameters, $V_{\rm rot,0}$ and $V_{\rm infall,0}$, to the model.

To compare with the observations, we convolve the modeled images with the beam of the observations and make the PV diagrams with the same PV cuts.
We calculate $\chi^2$ including both major and minor axes. 
Because CO and C$^{18}$O trace different velocity components, they are only used at specific ranges: $\mid V-V_{\rm lsr} \mid<1.6$ km s$^{-1}$ for C$^{18}$O and $\mid V-V_{\rm lsr} \mid>1.6$ km s$^{-1}$ for CO. 
In addition, the intensity ratio between CO and C$^{18}$O cannot be well fitted due to the unknown foreground absorption, spatial filtering, and probably the isotopic ratio. Thus, we include a scaling factor, $F_{\rm C^{18}O}$, for the C$^{18}$O emission as a free parameter in our fitting.
As a result, we find $\rho_0=5.5\times10^5$ cm$^{-3}$, $F_{\rm C^{18}O}=52$, $p=-2.2$, $\theta_{\rm flared}=35\arcdeg$, $V_{\rm rot,0}=1.6~\rm km~s^{-1}$, and $V_{\rm infall,0}=1.4~\rm km~s^{-1}$ in our best-fit with a reduced chi-squared, $\chi^2_{\rm r}=4.52$ (Figure \ref{fig:pv}).
This large $F_{\rm C^{18}O}$ and small $\rho_0$ imply that the observed CO intensity is unreasonably low.
This feature can be considered as a clue of CO depletion at the inner disk midplane that will be discussed later.
It is noteworthy that the flared structure with $\theta_{\rm flared}$ is required to reproduce the C$^{18}$O emission in the upper right and bottom left quadrants in the minor-axis PV diagram (Figure \ref{fig:pv}).

\subsection{CO channel map and CO-depletion model}
\label{sec:co_ch}
\begin{figure*}
\includegraphics[width=0.9\textwidth]{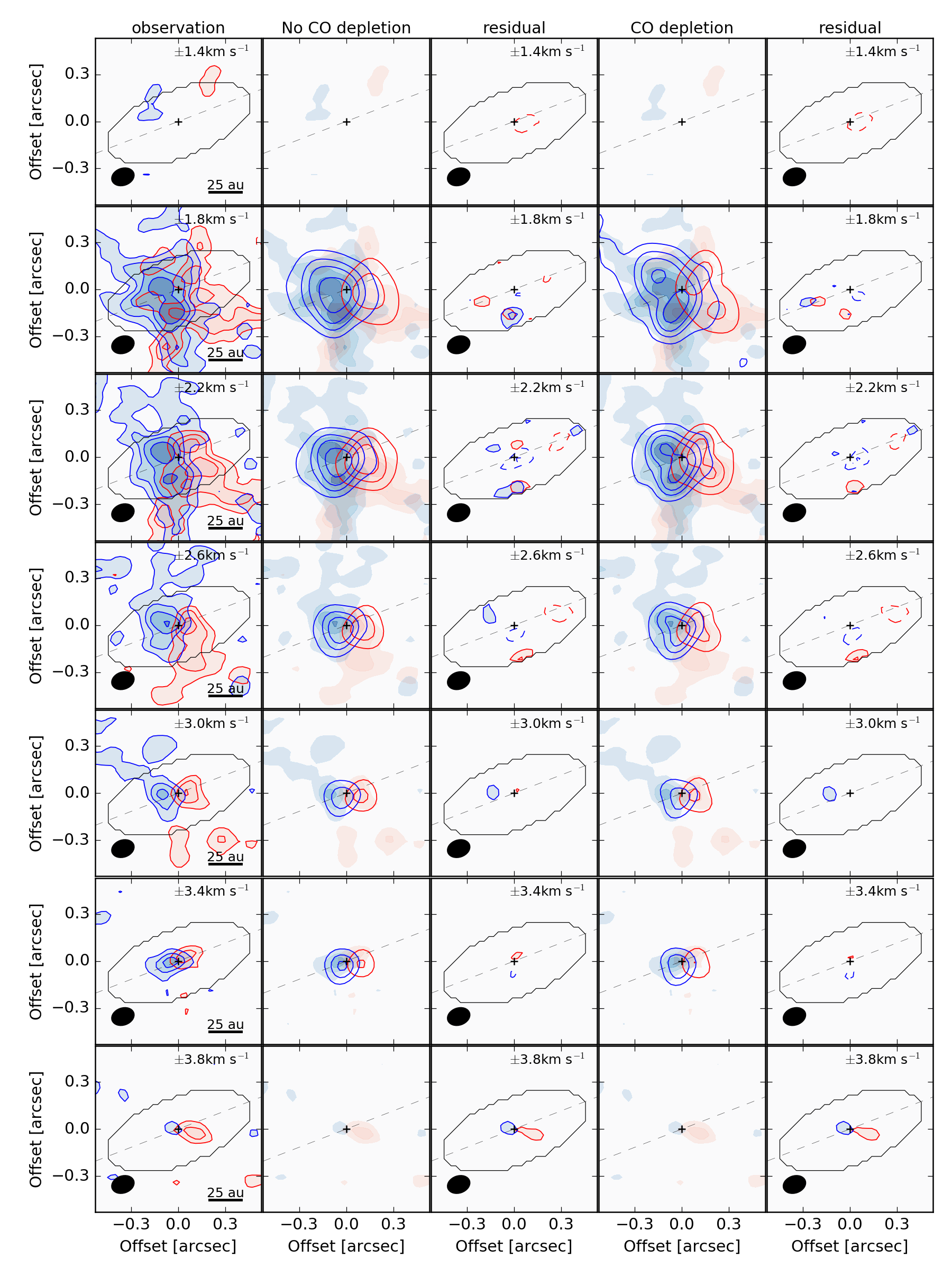}
\caption{Observed and modeled CO channel maps. The contours are plotted at 3, 5, 7, 10, and 15$\sigma$ with $\sigma$ of 3.3 mJy beam$^{-1}$. The first, second, and fourth columns show the contours of the observation, model without, and model with CO freeze out at the midplane, 
respectively, The filled contours in each of these panels show the observations for comparison.
The third and fifth columns show the residuals after subtracting the model of the previous column from the observations.
The elliptical grey lines indicate the mask used to fit the model.
The plus sign is the continuum source center and the dashed line represents its major axis.}
\label{fig:ch}
\end{figure*}

In addition to the PV diagrams, we compare the best-fit dynamic model from section \ref{sec:pv} with the observations in the CO channel maps (Figure \ref{fig:ch}).
The modeled images are processed through vis\_sample.
To reduce the effect from unknown foreground optical depths, the intensity of the model maps was scaled channel by channel. The scaling factor
of each channel was obtained by fitting the intensity of the model map with that of the observed one. In order to exclude the contamination of the outflow, the fitting
was applied to the region inside the elliptical mask shown in Figure \ref{fig:ch}. The scaling
factors have a mean value of 1.3 and a standard deviation of 0.4. This process does not significantly change the fitting nor the morphologies of the model.
Figure \ref{fig:ch} shows the resulting modeled images (the second column, no CO depletion) that generally fits the observation.

However, this model cannot reproduce the quadruple pattern seen in the low-velocity range between 1.8 and 2.2 km s$^{-1}$.
Considering a disk with an inclination angle of $65\arcdeg$, this pattern might originate from:
(1) absorption against the optically thick dust component in the disk/envelope midplane, (2) self-absorption of the optically thick gaseous CO, and (3) depletion of CO in the disk/envelope midplane.
Optically thick dust continuum emission at the disk midplane was reported in the Class 0 protostar HH212 \citep{le17a,le17b}.
However, it is unlikely the case for IRAS\,16253 because the 1.38 mm dust emission has a relatively small size, and such absorption is not seen in the high-velocity region, $\gtrsim2.6$ km s$^{-1}$. 
The second hypothesis is also unlikely because the far side (or bottom side) should be much fainter than the near side by being obscured.
In this case, we would expect to see highly asymmetric emission as the dust-continuum model in Section \ref{sec:con}.
The third possibility of CO depletion might be a reasonable explanation. 
CO freeze out at the disk midplane has been found in more evolved Class II sources \citep{qi13,qi15,sc16,pi18}, but it is unclear if such cold disks appear at the early stage \citep{va18}.

To mimic the CO depletion, we introduce a new free parameter, $f_{\rm snow}$ to our model;
the gas density at $z<f_{\rm snow}\times H(R)$ is set to zero, with $f_{\rm snow}$ between 0 (no CO depletion) and 1 (complete CO depletion).
As a result, we find the best-fit CO-depletion model for $f_{\rm snow}=0.55$ with $\chi^2_{\rm r}=1.78$ while $\chi^2_{\rm r}=1.92$ for $f_{\rm snow}=0$.
Figure \ref{fig:ch} shows the comparison between the models with and without CO depletion.
The model with CO depletion reproduces the quadruple pattern qualitatively well, though the CO peak positions are not perfectly matched and the difference between the $\chi^2_{\rm r}$ is small.

It is noteworthy that our model of the CO channel maps cannot distinguish CO depletion from the outflow-compressed gas or the outflow cavity wall irradiated by the central source.
These outflow features are commonly seen through CO \citep{ar06}. However, the velocity gradient of such gas components is usually along the outflow axes.
If it is the case, the surface layers require a significant higher excited state than the midplane while the latter likely has a higher density.

Another noticeable feature is seen in the channel maps: the orientation of the velocity gradient gradually converges into the disk major axis as the velocity increases.
This can be explained as an inner Keplerian disk surrounded by an infalling rotating envelope \citep{as15}. Thus, the high-velocity region could be dominated by rotation.
Assuming a pure rotation at the channels $\pm3.4~{\rm km~s^{-1}}$, we obtain a Keplerian velocity of $\approx3.7~{\rm km~s^{-1}}$ (deprojected) at a radius of $\approx7.8~{\rm au}$ by fitting a Gaussian to the CO emission in the channel maps.
This result however is in conflict with the 
assumption of free-fall for the infall motions (Section \ref{sec:pv}), and will be discussed in Section \ref{sec:dyn}.

\section{Implications and discussion}
\label{sec:discuss}
\subsection{grain growth at the Class 0 stage}
Grain growth is believed to be the signal of planet formation \citep{ap05}. 
This process has been studied in more evolved Class II circumstellar disks \citep{kw15,pi16}, but is not yet fully understood at earlier evolutionary stages.
Micron-sized grains could be formed in dense molecular clouds or cores \citep{pa10} and migrate into the later-formed protostellar disks.
Indications of large dust grains in inner envelopes or disks around Class 0/I objects are reported based on the opacity spectral indices, $\beta$, at submillimeter/millimeter wavelengths \citep{jo07,kw09,ch12}.
However, formation of (sub)mm-sized grains requires high density such as a disk midplane \citep{te14}.

Under the assumption of optically thin emission, we use the Gaussian fluxes in Table \ref{tab:gau} to derive a $\beta=1.4$ for the extended component and $\beta=0.9$ for the disk component assuming a dust temperature of 100 K and 30 K, respectively. These different indices imply that the dust size distribution has changed from the core to the disk; dust growth has started in the disk component. However, this analysis requires that the continuum emission at both wavelengths traces the same component. It is obviously not the case for the extended component because of the very different sizes. For the disk component, although the source structures are broadly consistent, it is still unclear whether the offset is real or not. 

Hundred micron-sized grains might have formed in IRAS\,16253 considering the offset between the continuum emission at 0.87 mm and 1.38 mm along the disk minor axis.
Our models suggest that an offset can originate from a large optical depth in an inclined disk, but it requires very different $\kappa$ between the observational wavelengths;
our best-fit model has $a_{\rm pow}=3.0$ and $a_{\rm max}=150~{\rm \mu m}$, resulting in an opacity ratio of $\sim$5.5 between these two wavelengths (Appendix A).
However, several caveats should be mentioned.
First, the offset of $0\farcs046$ is smaller than the beam sizes, $0\farcs15$ at 0.87 mm and $0\farcs1$ at 1.38 mm. 
Besides, the possibilities of a calibration issue or high proper motions have not yet been completely discarded.
Second, although very different $\kappa$ values at the two wavelengths do produce offsets in the images, our best-fit model does not fully reproduce the observed
large offset.
Multi-wavelength observations at higher angular resolution are required to provide better constraints for future detailed modeling.

\subsection{dynamics of the disk-envelope system}
\label{sec:dyn}
The infall and rotation velocities of IRAS\,16253 are estimated from the PV diagrams.
Assuming the infall motion is a free-fall motion, we estimate a central mass of $\sim$$0.028~M_\odot$. 
By equalizing the gravitational force and the centrifugal force from $V_{\rm rot}$, we found a centrifugal radius of $\sim$64 au.
However, our CO observations do not resolve any Keplerian rotation despite a small beam size of 15 au.
A possibility is that the true disk radius is only half the centrifugal radius, {\rm i.e.} the centrifugal barrier of $\sim$32 au \citep{sa14}.
An other explanation is that the centrifugal radius may be smaller if the infall velocity is smaller than the free-fall velocity.
Assuming the rotation dominates the velocity of $\pm3.4~{\rm km~s^{-1}}$ in Figure \ref{fig:ch}, we found a deprojected Keplerian velocity of 3.7 km s$^{-1}$ at a radius of 7.8 au.
We then estimate a mass of the central star of 0.12 $M_\odot$ (Section \ref{sec:co_ch}), such that the centrifugal radius is $\sim$$16~{\rm au}$ and centrifugal barrier is $\sim$$8~{\rm au}$.
In such a case, the infall velocity would be 50\% smaller than the free-fall velocity.
These two possibilities are not in conflict with each other, and the true system could be a mix of them.
Therefore, we speculate that the radius of the Keplerian disk is in between 8 and 32 au.

\subsection{The disk size and disk growth}
Only a small number of Class 0 protostellar disks have been kinematically identified while it is crucial to understand the disk formation. 
Our dynamical analysis suggests a Keplerian disk with a radius of $8-32$ au in the Class 0 source IRAS\,16253.
This result also broadly agrees with the size of the dusty disk $9-30$ au (Table \ref{tab:mod}).

Disk formation has not yet been fully understood, and two scenarios are proposed to explain the growth of the Keplerian disk radius: (1) ``early-start, slow-growth'' or (2) ``slow-start, rapid-growth'' \citep{ye17,le18}.
In the classical picture, i.e. ``slow-start, rapid-growth'', the growth of the disk radius in a non-magnetized collapsing core is:
\begin{equation}
r_{\rm kep}~({\rm au}) \sim 0.25 (\frac{\Omega}{10^{-14}~\rm rad~s^{-1}})^2 (\frac{a}{0.2~\rm km~s^{-1}}) (\frac{t}{10^{5}~\rm yr})^{3}
\end{equation}
where $\Omega$ is the initial cloud core rotation rate, $a$ is the sound speed, and $t$ is the time since the core collapse \citep{te84,be13}.
IRAS\,16253's cloud core rotation rate has been measured to be $3.5-4.1\rm~km~s^{-1}~pc^{-1}$, which is relatively small among 17 Class 0/I objects (\citealp[median: $8.1-10.7~\rm km~s^{-1}~pc^{-1}$,][]{to11}). In addition, it has the smallest N$_2$H$^+$ line width ($<0.2~{\rm km~s^{-1}}$) compared with other VeLLOs \citep{hs15,hs18}, implying a very small sound speed.
Such properties suggest that the disk in IRAS\,16253 might grow relatively slowly.
If we use $\Omega=\rm 1.8-2.1\times10^{-14}~rad~s^{-1}$ \citep{to11} and $a=0.14~\rm km~s^{-1}$ 
\citep{hs18}, it takes $2.5-3.7\times10^5~\rm yr$ to form a disk with $r_{\rm kep}=8-32~{\rm au}$. 
However, it is unrealistic because IRAS\,16253 is considered to be much younger due to its small mass fraction of the star+disk ($0.03-0.12~M_\odot$) to the core (\citealp[$0.2-0.8~M_\odot$:][]{ba10,st06,en08,to12b}); this fraction may suggest an age of $\lesssim0.5\times10^5~\rm yr$ in the nonmagnetic collapsing model \citep{yo05}.
Thus, the disk of IRAS\,16253 seems to favor the ``early-start, slow-growth'' scenario that is supported by the analysis of
the properties of the Class 0 source HH211 \citep{le18}.

A possible explanation for this scenario might be passing through a rotating first hydrostatic core (FHSC) with a size of a few au and a lifetime of a few thousand years \citep{la69}. 
Theoretical models suggest that a rapidly rotating FHSC may directly evolve into a circumstellar disk after the collapse \citep{ba11,ma11}.
However, given IRAS\,16253's small cloud rotation rate, the disk size after the collapse should still be as small as a few au.
High-angular-resolution observations are needed to resolve the size of the Keplerian disk and examine this disk formation process.


\subsection{freeze out of CO in the protostellar disk}
CO depletion is considered as the most plausible explanation for the quadruple pattern in the channel maps although the other hypotheses cannot be completely ruled out.
The freeze out of CO would suggest a temperature below 20 K, the CO sublimation temperature, in the midplane of the disk or the inner envelope.
This kind of ``cold disk'' had not yet been found around protostars at an early embedded stage. 
\citet{ha15} found that, unlike more evolved Class II disks, embedded disks can be heated by viscous accretion and stay warm due to the inefficient radiative cooling in the optically thick envelope. 
Such a picture is confirmed toward a borderline Class 0/I protostar, L1527; it shows gaseous CO throughout the disk without detection of N$_2$D$^+$ which is abundant when CO is frozen out \citep{va18}.
The CO freeze out in IRAS\,16253 may result from its unique low internal luminosity as a VeLLO, providing low radiative heating.
In addition, the disk/envelope midplane could be shielded or self-shielded from heating by the central protostar given the optically thick disk (Section \ref{sec:con}), as seen in VLA1623 \citep{mu15}.


It is also noteworthy that IRAS\,16253 has experienced a past accretion burst that temporally enhanced the protostellar luminosity and sublimated CO within a radius of $\sim$1250 au \citep{hs18}. 
The CO freeze-out timescale is a function of dust temperature, $T_{\rm g}$, and gas density, $n_{\rm H_2}$,
\begin{equation}
\tau_{\rm fr}=1\times10^4 \sqrt{\frac{T_{\rm g}}{10 K}}\frac{10^6~{\rm cm^{-3}}}{n_{\rm H_2}}~{\rm yr}
\end{equation}
\citep{vi12}.
Thus, if the gas density in the depletion region is $>10^7~{\rm cm^3}$, it requires $<1000$ yr for CO to refreeze out.
In addition, since the envelope is infalling, the gas might have migrated from the outer region into the inner region.
For example, the CO frozen-out gas at $r\sim15$ au could have migrated from $r\sim100$ au assuming an infalling velocity of $\rm 1~km~s^{-1}$ in 500 yr.
Therefore, if the density structure is resolved, a chemical-dynamical model would allow us to measure the time since the last accretion burst.

\section{Summary}
We present ALMA long baseline observations toward the Class 0 source IRAS\,16253-2429. We summarize our results in the following:
\begin{enumerate}
\item A compact source is detected from the continuum emission at both 0.87 mm and 1.38 mm.
An offset of $\sim$46 mas is found between the continuum emission peaks at 0.87 mm and 1.38 mm although a calibration issue cannot be completely ruled out. This offset along the outflow axis could originate from the different optical depths in an inclined disk. 
However, it requires a very large $\kappa$ ratio between these two wavelengths.
The largest ratio we can reproduce is $\sim5.7$ at $a_{\rm max}=150~{\rm \mu m}$ and $a_{\rm pow}=3.0$, for which $a_{\rm pow}=3.0$ is the minimum in our parameter space.
Our model does generate an offset of $10-20$ mas, but it is still smaller than the observed value.
\item Rotation and infall motions are detected through CO and C$^{18}$O ($2-1$) emission toward the disk-envelope system. Assuming the infall motion is free fall, we estimate the central stellar mass to be $\approx0.03~M_\odot$. However, the rotation motion implies a mass of $\sim0.12~M_\odot$, and in this case, the infall velocity is reduced by $\sim$50\% from free fall.
Further observations are required to test these two possibilities, and help to answer if IRAS\,16253 will form a brown dwarf ($<0.08 ~M_\odot$) or a normal low-mass star in the future.
\item The best-fit dynamical model has a centrifugal radius of $\sim$$64~{\rm au}$, but the Keplerian rotation is not resolved by the current resolution of 15 au. 
Deriving from the rotation dominated region, we estimate a centrifugal radius of 16 au.
Together with the size from the dust continuum, we speculate that a Keplerian disk is present with a radius of $8-32$ au in IRAS\,16253.
\item The presumed disk radius, $8-32$ au, is much larger than that derived from the classical non-magnetized collapsing model given the small cloud core rotation rate and sound speed in IRAS\,16253.
Therefore, the circumstellar disk in IRAS\,16253 may have directly evolved from a rotating first hydrostatic core, as suggested by theoretical models.
\item The quadruple pattern in the CO channel maps at low velocities could be explained by freeze out of CO in the disk midplane.
The presence of such a ``cold disk'' may result from the faint luminosity of the protostar. Besides, the dense inner disk, as indicated by the continuum images, might shield the outer region or be self-shielding from the central heating source.
\end{enumerate}

\acknowledgments
We are thankful for the anonymous referee for many insightful comments that help to improve this paper significantly.
The authors thank Dr. Hsi-Wei Yen, for providing valuable discussions.
We thank Dr. Attila Juhasz for helping in running the RADMC-3D code.
We are thankful for the help from ALMA Regional Center in Taiwan.
This paper makes use of the following ALMA data: ADS/JAO.ALMA\#2016.1.00598.S, 2013.1.00879.S, and 2015.1.00741.S. ALMA is a partnership of ESO (representing its member states), NSF (USA) and NINS (Japan), together with NRC (Canada), MoST and ASIAA (Taiwan), and KASI (Republic of Korea), in cooperation with the Republic of Chile. The Joint ALMA Observatory is operated by ESO, AUI/NRAO and NAOJ.
N.H. acknowledges a grant from the Ministry of Science and Technology (MoST) of Taiwan (MoST 107-2119-M-001-029).
SPL acknowledges support from the Ministry of Science and Technology of Taiwan with Grant MOST 106-2119-M-007-021-MY3.

\setcounter{figure}{0}
\setcounter{table}{0}
\renewcommand{\thefigure}{A\arabic{figure}}
\section*{Appendix A. Opacity model}
\label{app:kappa}
\begin{figure}
\includegraphics[width=0.45\textwidth]{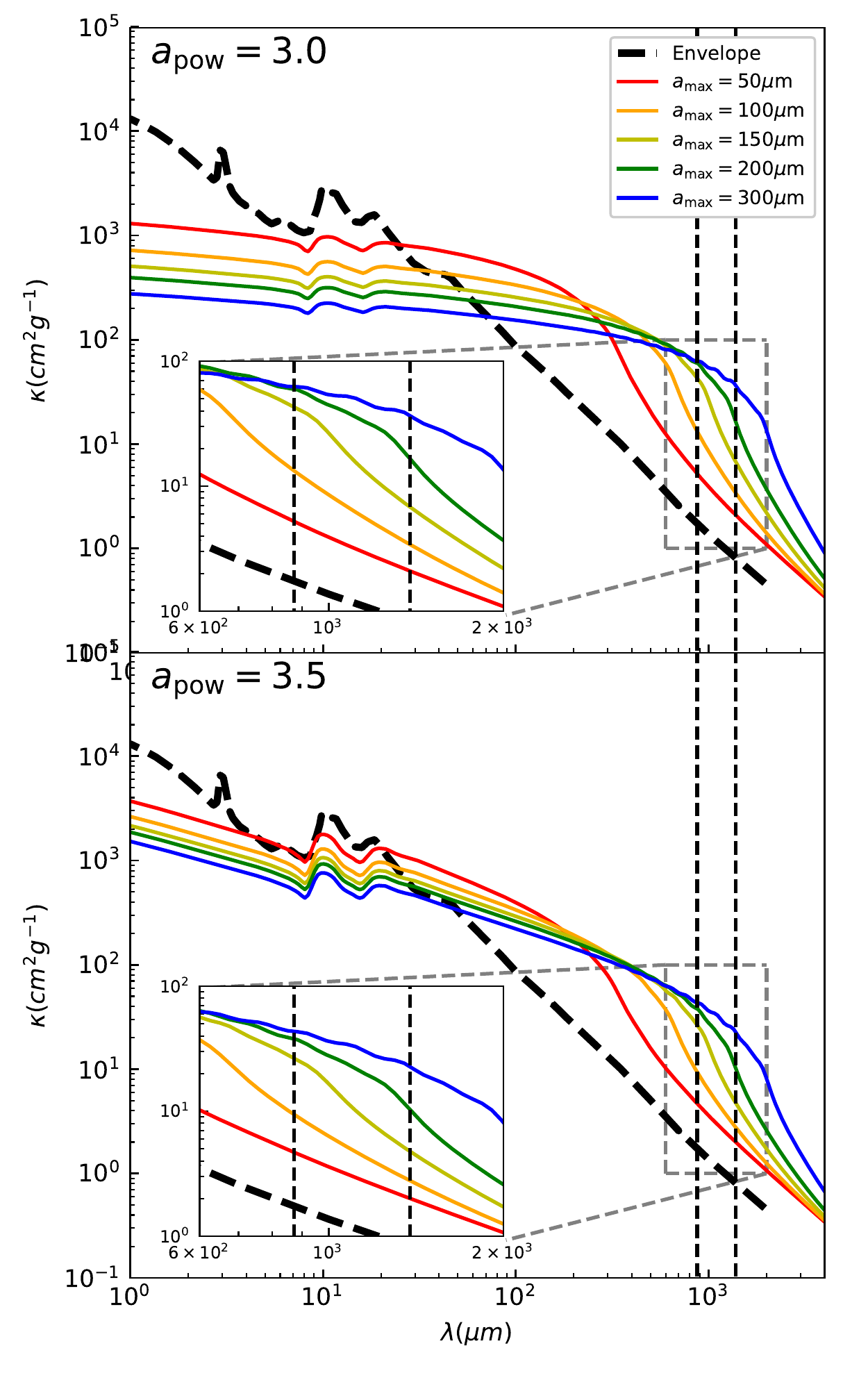}
\caption{Dust opacity as a function of wavelength for the dust size distribution with a power-law index of 3.0 (top) and 3.5 (bottom). The colored solid lines show models with different maxinum dust sizes. The thick dashed line represents the model with thin ice grain coagulated at a density of $10^6~{\rm cm}^3$ \citep{os94} as a reference. The two vertical dashed lines indicate the observed wavelengths, 0.87 and 1.38 mm, in this paper.}
\label{fig:kap}
\end{figure}

The dust opacity used in the continuum model (section \ref{sec:con}) was obtained from the DIANA Opacity Tool \citep{wa16}. The DIANA opacity tool computes fast models of the dust opacity $\kappa$ as a function of wavelength. 
Dust opacities for absorption $\kappa_{\rm abs}$ and scattered $\kappa_{\rm sca}$ are derived and are used in the RADMC-3D code.
Figure \ref{fig:kap} shows the summation of $\kappa_{\rm abs}$ and $\kappa_{\rm sca}$ with different maximum sizes of dust. An extreme case with the maximum opacity ratio between 0.87 mm and 1.38 mm is found ($\kappa_{0.87 \rm mm}/\kappa_{1.38 \rm mm}\approx5.5$) when $a_{\rm max}=150~\mu$m and $a_{\rm pow}=3.0$. 
We note that this ratio corresponds to an index $\beta\approx4.2$ which is even larger than that in the interstellar medium. 
However, this index was measured between 0.87 mm and 1.38 mm. It may not be representative of the index over a broader wavelength
range.

\newpage

\bibliographystyle{aasjournal}


\end{document}